\documentclass[aps,showpacs]{revtex4}
\usepackage{graphicx}
\usepackage{amsfonts}
\def\beqn{\begin{equation}}
\def\eeqn{\end{equation}}
\def\beqnar{\begin{eqnarray}}
\def\eeqnar{\end{eqnarray}}

\def\ba{\begin{array}}
\def\ea{\end{array}}

\newcommand{\mket}[1]{\vert{#1}\rangle}
\newcommand{\mbra}[1]{\langle{#1}\vert}

\newcommand{\ignore}[1]{}

\newcommand{\eqn}[1]{(\ref{#1})}
\newcommand{\eqnRef}[1]{Eq. (\ref{#1})}
\newcommand{\figRef}[1]{Fig.~\ref{#1}}

\begin{document}

\title{Robust Control of Quantum Information}

\author{Marco A. Pravia}
\affiliation{Department of Nuclear Engineering, Massachusetts Institute of Technology, Cambridge, MA 02139}
\affiliation{Department of Computer Science, University of Puerto Rico, Rio Piedras, PR 00931}
\author{Nicolas Boulant, Joseph Emerson}
\affiliation{Department of Nuclear Engineering, Massachusetts Institute of Technology, Cambridge, MA 02139}
\author{Amro Farid}
\affiliation{Department of Mechanical Engineering, Massachusetts Institute of Technology, Cambridge, MA 02139}
\author{Evan M. Fortunato, Timothy F. Havel, David G. Cory}
\affiliation{Department of Nuclear Engineering, Massachusetts Institute of Technology, Cambridge, MA 02139}

\vspace*{.35in}

\begin{abstract}
Errors in the control of quantum systems may be classified as unitary, decoherent and incoherent.
Unitary errors are systematic, and result in a density matrix that differs from the desired one by a unitary
operation. Decoherent errors correspond to general completely positive superoperators, and can only be
corrected using methods such as quantum error correction. Incoherent errors can also be described,
on average, by completely positive superoperators, but can nevertheless be corrected by the 
application of a locally unitary operation that ``refocuses'' them. 
They are due to reproducible
spatial or temporal variations in the system's Hamiltonian, so that information on the variations
is encoded in the system's spatiotemporal state and can be used to correct them. In this paper liquid-state
nuclear magnetic resonance (NMR) is used to demonstrate that such refocusing effects can be built directly
into the control fields, where the incoherence arises from spatial inhomogeneities in the quantizing static
magnetic field as well as the radio-frequency control fields themselves. Using perturbation theory, it is
further shown that the eigenvalue spectrum of the completely positive superoperator exhibits a 
characteristic spread that contains information on the Hamiltonians' underlying distribution.

\end{abstract}


\maketitle

\newpage

\section{Introduction}

Methods of controlling quantum systems \cite{Rabi,Waugh, Haught-Control,Warren-Control, Warren-Review}
are needed to direct the course of chemical reactions \cite{GordonR97,Rice-Active-Control-Review}, 
to determine molecular structure \cite{ErnstBook}, 
and to achieve quantum information processing \cite{NielsenBook}.  The common goal 
is to preserve or manipulate a quantum system so that the effective evolution
over a control sequence is precisely the desired process. 
The causes of unwanted dynamics include 
irreversible couplings to the environment (decoherence), static or slowly varying fluctuations 
in the system's Hamiltonian (incoherence), and systematic unitary errors.  
Here, we examine a common class of experimental imperfections 
characterized by a spatially or temporally ``incoherent variation'' in the system's Hamiltonian. 
In addition, we describe a method for mitigating this class of
errors, and we report an experimental demonstration using liquid-state nuclear
magnetic resonance (NMR) techniques \cite{ErnstBook}.

Experimental limitations present an important set of challenges to achieving
precise control over quantum systems.  Incoherent errors, which are often
present in experiments, may be distinguished from decoherent
because they are in principle refocusable. In the case of a spatially distributed ensemble
interacting with a field, the field amplitude and direction can vary over the ensemble
so that the ensemble average dynamics becomes a convex sum of spatially distinct
unitary processes. The ensemble's dynamics will in
general appear non-unitary, but as long as the correlation between 
the spins' location and the strength/direction
of the field remains unchanged the dispersion in the dynamics can be refocused,
at least in principle.

Spatially incoherent errors have been a recurring topic of interest in the field of NMR,
where they arise as inhomogeneities in the static and radio-frequency (RF) fields involved.
The spatially incoherent evolution caused by the inhomogeneities dephases
the spins in the NMR ensemble, attenuating and rotating the final 
state away from the desired state.  Past methods of refocusing incoherent evolution 
have been directed mainly towards achieving rotations of specific states rather than implementing 
a desired unitary operation on all possible states. Examples include ``composite pulses'' 
\cite{LevittF79, Levitt82, Tycko83, Shaka-DualComp, Levitt86}, which have played an important role 
in creating the robust RF pulse sequences used for spin inversion, spin excitation, 
and decoupling.  Adiabatic pulses have also 
been used to overcome the problem of RF inhomogeneity \cite{BaumTP85, SilverJH85}.
In addition, average Hamiltonian methods \cite{Waugh} have been applied to create 
pulse sequences that depend mostly on phase changes in the RF fields to control the spins, 
making the RF field's amplitude a less important experimental 
parameter \cite{Cory-RFSelection}.  Most of this work has been
devoted to designing robust one-spin operations, but more recently 
the work has been extended, in the context of quantum information processing, to 
include two-spin operations \cite{JonesRobust}.

Here, we extend our previously described method \cite{Fortunato-Control} for
creating RF control gates in liquid-state NMR quantum 
information processing experiments \cite{Cory-PP, Chuang-PP, Jones-NMRQIP, Cory-Overview}.
The gates use RF fields that strongly 
modulate a system's dynamics so as to refocus all the undesired evolutions and 
to achieve a specific desired unitary transformation with high precision.
The method makes use of numerical optimization and complete knowledge of the 
system's internal Hamiltonian to find a modulation of the RF field such that the
effective Hamiltonian over the period for which it is applied is precisely
the desired transformation.  
In this paper, we extend this method so as to also use knowledge
of the incoherent distribution of errors in an effort to design
pulses that are robust over the entire distribution and thus
yield the desired transformation with even higher fidelity.  

\section{Modeling, Measurement and Analysis of Incoherent Processes} \label{ControlParamSearchProc}

In quantum information processing, the central experimental goal is to
efficiently generate any desired unitary operation on a quantum system. 
Unitary operations are realized by manipulating externally-controlled parameters in 
the system's Hamiltonian.  The problem of finding the appropriate external parameters,
however, is an inverse problem, and analytical solutions are not available except for 
the simplest cases. As described below, we can utilize the more easily solved
forward problem in a numerical search for a control field modulation that
solves the inverse problem, at least in small Hilbert spaces.
The density matrix evolution under an incoherent process appears to evolve in a
larger space with a continuum of parameters, denoted by $\overrightarrow{r}$,
describing the variations over the ensemble.
In NMR, $\overrightarrow{r}$ would be the spatial location of the individual molecules. 
Every measurement is an integral over $\overrightarrow{r}$, so assuming that 
the local density matrix $\rho(\overrightarrow{r})$ is uncorrelated with the 
local unitary operation $U(\overrightarrow{r})$, the ensemble-average density matrix
$\rho_{in} = \int  \rho(\overrightarrow{r}) d\overrightarrow{r}$ 
evolves in Hilbert space as
\begin{equation} \label{contKraussForm}
\rho_{out}=\int U(\overrightarrow{r})\rho_{in} U^{\dagger}(\overrightarrow{r})d\overrightarrow{r}
\label{IncEvol}
\end{equation}
The variation of the operator $U(\overrightarrow{r})$ is tied to the variation
of a field in the corresponding Hamiltonian.
The evolution can be expressed in matrix form as a superoperator
acting on Liouville space, i.e.
\begin{equation}
\mket{\rho_{out}}=\int \left( \overline{U}(\overrightarrow{r})\otimes
U(\overrightarrow{r}) \right) d\overrightarrow{r} \mket{\rho_{in}}
\end{equation}
where $\mket{\rho}$ is the columnized density matrix (obtained by stacking its columns
on top of each other), $\overline{U}$ is the complex conjugate of $U$ and ``$\otimes$''
is the Kronecker product of the matrices \cite{ErnstBook,Havel-Superop}.
In this form, it is clear that the input state $\mket{\rho_{in}}$ is transformed by
the superoperator
\begin{equation}
S=\int \left( \overline{U}(\overrightarrow{r})\otimes U(\overrightarrow{r}) \right) d\overrightarrow{r}
\end{equation}
to produce the final state $\mket{\rho_{out}}$.
So although every element of the ensemble evolves unitarily,
the ensemble-average density matrix undergoes non-unitary dynamics.

\subsection{Incoherent Processes in NMR Spectroscopy}

In the specific case of liquid-state NMR, the task of quantum control is
to determine the RF pulse sequence that modulates the internal spin
Hamiltonian of every molecule so as to generate a desired evolution.  The 
homonuclear internal Hamiltonian for a molecule containing $N$ spin $1/2$ 
nuclei is
\begin{equation}
H_{int}=\sum_{k=1}^N -\gamma_k(1-\sigma_z^k)B_0(\overrightarrow{r}) I^k_z + 2\pi\sum_{j>k}^N\sum_{k=1}^N J_{kj} \\
I^{k}{\cdot}I^{j},
\end{equation}
where $-\gamma_k(1-\sigma_z^k)B_0(\overrightarrow{r})$ represents the chemical shift frequency
of the $k$th spin ($\gamma$ is the gyromagnetic ratio and $\sigma_z$ is the shielding constant), 
$J_{kj}$ is the coupling constant between spins $k$ and $j$ and $I_i$ denotes the i axis spin angular momentum operator.
The chemical shifts are functions of space because the
main magnetic field is never perfectly homogeneous througout the sample.
The scalar coupling constants, however, are independent of location since
they depend only on molecular bonding.

The corresponding experimentally-controlled RF Hamiltonian is
\begin{equation}
H_{ext}(t)=\sum_{k=1}^N -\gamma_k f(\overrightarrow{r})B_{RF}(t)e^{-i\phi(t)I_z^k}I^k_x e^{i\phi(t)I_z^k},
\end{equation}
where the time-dependent functions $B_{RF}(t)$ and $\phi(t)$ specify the applied RF control field, while
$f(\overrightarrow{r})$ reflects the RF field strength distribution over the sample.  Here, 
we assume that only phase changes in time are important and drop the spatial dependence of $\phi$.

Excluding decoherence, the evolution generated by the above Hamiltonians between time $0$ and $\tau_f$ is
\beqn \label{Unet}
U_{gate}(\overrightarrow{r}) = T \text{exp} \left( -i \int^{\tau_f}_{0} \left[ H_{int}(\overrightarrow{r})
        +H_{ext}(\overrightarrow{r}, t)\right] dt \right),
\eeqn
where $T$ is the Dyson time-ordering operator.  The goal
is to determine a function $B_{RF}(t)$
and $\phi(t)$ that results in a net evolution that at every location is close to the
desired transformation.  In general, this inverse problem is difficult
to solve but the forward problem of evaluating $U_{gate}$ is readily
solved numerically for small spin systems.

To efficiently calculate $U_{gate}(\overrightarrow{r})$, several simplifications
can be made.  First, the use of shim controls allows the static magnetic field
homogeneity to be made less than 1 part in $10^8$ over the sample volume, meaning that the resonance 
frequencies of magnetically equivalent spins differ
by the same amount.  As a result, for now we will drop the spatial dependence of $B_0$;
the major problem is the RF field variations in $H_{ext}$.
We have shown previously that in the fully coherent case strongly modulating 
RF pulses with piecewise constant RF amplitudes provide an easily computed
modulation sequence with sufficient control over the spins \cite{Fortunato-Control}.
The evolution of a single pulse
with constant (in time) RF power $B_{RF}$,
phase $\phi$, frequency $\nu_{RF}$, and of duration $\tau$,
can be solved with one frame transformation and a single
diagonalization in that frame.  This simplification allows the
net propagator of a train of $M$ such square pulses to be
written as
\begin{equation}
U_{gate}= \\
   \prod_{m=1}^M U^{-1}_z(\nu_{RF,m}, \tau_m)  \\
   \text{exp}\left[-i H_{eff}^m(B_{RF,m},\nu_{RF,m},\phi_m) \tau_m\right]
\label{Usquare}
\end{equation}
where $U^{-1}_z(\nu_{RF,m}, \tau_m)$ executes the rotating-frame
transformation of the $m$th period and $H_{eff}$ is the effective,
{\em time-independent} Hamiltonian in the new frame of reference \cite{ErnstBook}.
Using the standard simplex search algorithm, a set of parameters
which match $U_{gate}$ to a desired transformation can often be
obtained \cite{Fortunato-Control}.  
In the coherent case, we found the dynamics of these pulses
to be very rich, admitting many different strongly-modulating
pulses for a given ideal unitary gate. Here we extend this solution
to the case of incoherent processes in the control Hamiltonian.

Radio-frequency power inhomogeneity was incorporated into the parameter search
by tabulating a discrete histogram of the RF power variations.  This distribution
of RF power defines a Kraus operator sum \cite{NielsenBook,Havel-Superop}
instead of the unitary transformation of \eqnRef{Usquare} , namely
\beqn \label{inhomoKraussOps}
\rho_{out} \;=\; {\sum}_k\, A_k\, \rho_{in} A_k^\dag \;,
\eeqn
where $A_k = \sqrt{p_k} U_k$ and $p_k$ is the fraction of the ensemble that undergoes
a unitary transformation $U_k$.  The operators $U_k$ were evaluated using \eqnRef{Usquare},
and they differ only in the RF amplitudes involved.  In NMR spectrometers the 
frequency and duration of the pulse do not vary as a function of position, and although 
the absolute phase does, this phase is unobservable here since the same RF coil is 
used for both transmission and reception.  

\begin{figure}
\centerline{\includegraphics[width=0.5\textwidth]{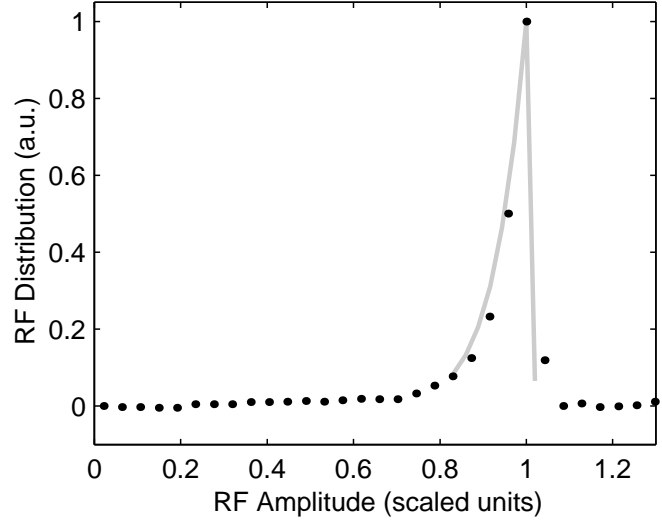}}
 \vspace{.25cm}
 \caption{{\bf Radio-frequency inhomogeneity profile.} The RF inhomogeneity in the carbon channel
was measured using a spin nutation experiment.  The resulting decaying signal was
Fourier transformed to distill the various RF nutation frequencies present in 
the sample.  The dotted line is the plot of the Fourier transformation, and
it is the measured RF inhomogeneity profile.  The solid gray line is the 
profile that was used to design pulses compensated for RF inhomogeneity, 
and it was extracted from the
measured profile.  }
 \label{InhomoProf}
\end{figure}

The RF inhomogeneity of our experimental setup was measured using
a spin nutation experiment in which the transverse magnetization
of the spin system was measured after on-resonance pulses of
increasing duration. The power spectrum of nutation frequencies (\figRef{InhomoProf})
is a direct measure of the distribution of RF field strengths over the sample.

The inclusion of RF inhomogeneity in the calculation of the fidelity,
as outlined in section \ref{ControlParamSearchProc}B, increases
the required computational resources per iteration by a factor equal to the
number of intervals used to define the distribution.  Hence for the 
design of the pulses here only $9$ intervals were used, and this distribution
is superimposed on the experimental measurement in \figRef{InhomoProf} (gray line).

\subsection{Metrics for the precision of coherent control} \label{MetricsOfControl}
Two useful metrics of coherent control are the state
correlation and the gate fidelity \cite{Fortunato-Control}. 
The correlation measures the closeness of two density matrices
$\rho_{ideal}$ and $\rho_{out}$ resulting from the same input
state $\rho_{in}$ but evolved under different propagators, 
one of which is an ideal gate $U_{ideal}$ such that
\begin{equation}\label{eq:rhotheo}
\rho_{ideal}=U_{ideal}\rho_{in}U_{ideal}^{\dag}
\end{equation}
%\beqn 
%\label{rho_th}
%\rho_{in} \stackrel{U_{ideal}}{\longrightarrow} \rho_{ideal}.  
%\eeqn
while the other is the simulated non-unitary transformation 
given by \eqnRef{inhomoKraussOps}
\begin{equation}\label{eq:incoherent}
\rho_{out}=\sum_kA_k\rho_{in}A_k^{\dag}
\end{equation}.
%\beqn 
%\label{rho_out}
%\rho_{in} \stackrel{\left\{A_k\right\}}{\longrightarrow} \rho_{out}.  
%\eeqn 
The non-unitary transformation is trace-preserving and completely-positive.
The correlation between the traceless parts $\hat\rho_{ideal}$ and $\hat\rho_{out}$
of the density matrices $\rho_{ideal}$ and $\rho_{out}$ is 
\beqn \label{StateFid}
C(\hat\rho_{ideal}, \hat\rho_{out}) = \frac{trace(\hat\rho_{ideal}~\hat\rho_{out})}
	{\sqrt{trace(\hat\rho_{ideal}^2) trace(\hat\rho_{out}^2)}},
\eeqn 
which varies from $-1$ to $1$.
The correlation reports only similarities
in the ``direction'' between two density matrices.
To account for the loss of information
due to non-unitary operations, an attenuation factor is
inserted, leading to the attenuated correlation, which 
is defined by
\begin{eqnarray} \label{AttnStateFid}
C_A(\hat\rho_{ideal}, \hat\rho_{out}) = C(\hat\rho_{ideal}, \hat\rho_{out}) 
	\sqrt{\frac{trace(\hat\rho_{out}^2)}{trace(\hat\rho_{in}^2)}}\\
	= \frac{trace(\hat\rho_{ideal}~\hat\rho_{out})}
	{\sqrt{trace(\hat\rho_{ideal}^2) trace(\hat\rho_{in}^2)}}.
\end{eqnarray}
The attenuation term quantifies the amount of information lost due
to a decrease in the magnitude of $\hat\rho_{out}$ relative to $\hat\rho_{in}$.

The gate fidelity $F$ is a measure of the precision of an operation
\cite{Schumacher-Channels, Fortunato-Control, Fortunato-DFS}
\beqn
\label{GateFidelity} 
F = \overline{C_A(\hat\rho_{ideal}, \hat\rho_{out})}, 
\eeqn 
where the overline notation $\overline{C_A}$ represents the average 
attenuated correlation over a complete basis of orthonormal Hermitian
matrices $\rho_j$ ({\it i.e.}, $trace[\rho_j\rho_k]=\delta_{jk}$).
In the case of incoherent processes, it is more convenient to use the equivalent expression for $F$ directly
in terms of $U_{ideal}$ and the Kraus operators $\left\{A_k\right\}$, namely
\beqn \label{ReducedF}
F=\frac{1}{2^{2N}}\sum_k\left|Tr(U_{ideal}^{\dag}A_{k})\right|^2
\eeqn
where $N$ is the number of qubits,
which can be evaluated more rapidly than the form in \eqnRef{GateFidelity}.

\subsection{Exploring the Achievable Fidelities} \label{MaxFids}

\begin{figure}
\centerline{\includegraphics[width=8.5cm]{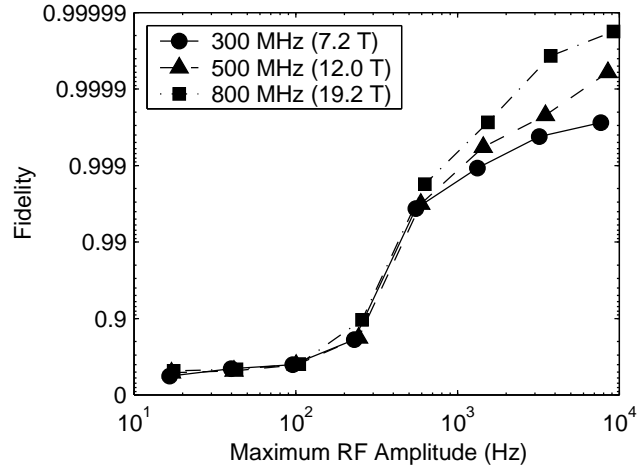}}
	\vspace{.25cm} 
	\caption{{\bf Numerical exploration of achievable fidelities.}  
The plot shows the maximum fidelities found for a $\pi / 2 ]_x^2$ alanine 
pulse as allowed RF amplitude and magnetic field strength $B_0$ were varied.  The 
three lines correspond to the magnetic field strengths explored, as 
denoted by the legend. }
   \label{AchievableFidsMax}
\end{figure}

In this subsection, we briefly revisit the coherent case.
The ideal single-spin gate fidelities of the pulses previously
achieved values over 0.999 in numerical simulations \cite{Fortunato-Control}.
Two important experimental parameters place restrictions on the
achievable fidelities, the main field strength $B_0$, 
which determines how well the individual spins are resolved,
and the maximum allowed RF power, which limits the modulation rate.
To learn how changing these parameters affects the fidelity of a gate,
we evaluted the fidelity for a $\pi/2]_x^2$ alanine pulse as a function
of allowed main field strength and RF amplitude.   The results are 
illustrated in \figRef{AchievableFidsMax}.

The three curves represent the results for each of three
static magnetic field strengths tested.  Each point represents
the highest fidelity found by the pulse search method out of  
several attempts to find the gate at that magnetic field and
RF power.  The stronger static fields
enabled pulses with higher fidelities because the stronger fields cause
the differences among the spins' frequencies to increase,
allowing more lattitude for addressability and control.
The fidelities also increased monotonically
with the maximum allowed RF power.
At low powers, the RF control is insufficient
to average out the internal Hamiltonian, resulting in low
fidelities, while at high RF power, the strength of the
RF dominates the internal Hamiltonian and hence allows
the dynamics to be more precisely controlled.

It is important to emphasize
that the maximum achievable fidelities of \figRef{AchievableFidsMax}
represent the best gates achieved using the current pulse
parameterization and search method and the available computational resources.  
These results do not preclude other methods and search strategies 
from finding higher fidelity gates.  In addition, the analysis explores
an optimistic scenario in which natural decoherence and experimental
imperfections, including RF field inhomogeneity, are ignored.  Nevertheless, 
expect these trends will also hold for the ``self-refocusing'' pulses 
developed in this paper.

\section{Designing Pulses That Compensate for RF Inhomogeneity} \label{robustnessSection}

%\section{Simulations of RF Inhomogeneity}
%   . Effect of increasing inhomogenity on uncompensated pulse\newline
%   . Adjusting pulse finder to compensate\newline
%   . effect of increasing inhomogenieity on compensated pulse\newline
%   . comparison of full set of pulses\newline
%\vspace{.25in}
%\subsection{Robustness Against RF Inhomogeneity} 

The spatial incoherence of the RF amplitude, or RF inhomogeneity, in NMR is a consequence
of the need for high sensitivity,
which necessitates wrapping the RF coil tightly around the sample.
As a result, different parts of the  NMR sample feel unequal RF fields,
causing a dephasing of the spins 
that attenuates the signal and introduces errors in the rotations of 
the spins.  The problem can be avoided by using smaller 
parts of the sample (and thus less signal).
An alternative, however, is to take advantage of the reproducibility of the
field strength distribution so as to design gates that are insensitive to these errors.  

\begin{figure}
\centerline{\includegraphics{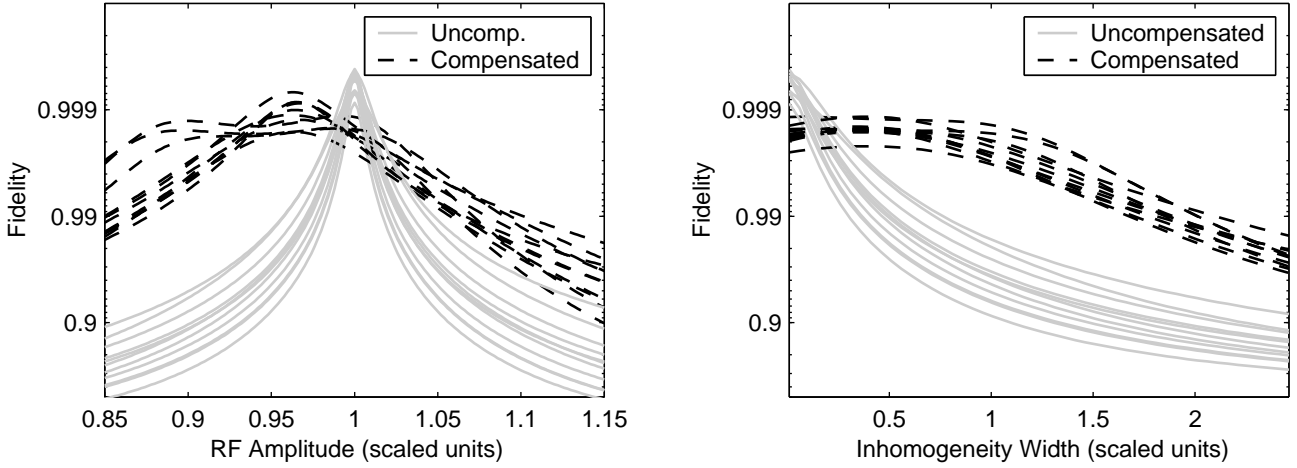}}
	\vspace{.25cm} 
	\caption{{\bf Simulations of compensated and uncompensated pulses as a function of
radio-frequency strengths and distribution widths.}  The dashed lines correspond
to compensated pulses, while the solid gray lines denote uncompensated
pulses. The left plot shows how the compensated pulses maintain high fidelities
even when the RF strength is scaled from the ideal value.  The plot on
the right simulates the same pulses as a function of the scaled width 
of the RF inhomogeneity profile.  These results demonstrate
the improved fidelity of the compensated pulses for all but the narrowest
RF distributions.  At the small widths, the RF profile would no
longer be inhomogeneous, eliminating the need for the compensated
gates.}	
   \label{robPlots}
\end{figure}

We used the methods of section \ref{ControlParamSearchProc} to search for both
compensated and uncompensated modulation sequences for 11 single qubit transformations.
The calculations were performed for the three-spin system consisting of the 
$^{13}C$-nuclei in isotopically-enriched alanine.  \figRef{robPlots} summarizes 
the simulated fidelities for the resulting gates.

Each fidelity point on the left of \figRef{robPlots} was calculated for a unitary
gate having a single RF field strength.  Each curve traces the gate
fidelity as a function of the deviation from the ideal RF field strength. 
The plot shows that the compensated pulses
are significantly less sensitive to changes in the RF modulation strength.
The uncompensated pulses, however, have the overall highest fidelities when 
the RF amplitude is at its optimum value.
This fact is confirmed in the more realistic situation where a spectrum of RF frequencies 
is present.  The plot on the right of \figRef{robPlots} graphs the pulse
fidelities for the non-unitary transformations generated with RF distributions
of varying widths.  The distributions used were stretched or 
narrowed versions of the measured RF profile (\figRef{InhomoProf}), and the 
widths have been rescaled such that a value of $1.0$ corresponds to the 
experimentally measured RF profile.  

\section{Eigenvalue Spectra of Superoperators}

In this section, we present a numerical study of the action of both compensated and uncompensated gates
by looking at the eigenvalue spectrum of the superoperators of the gates.  We will show
how the eigenvalues of the superoperators can serve as a useful and convenient tool for
extracting information about the imperfections in the implementations of unitary transformations.
In particular, we shall describe features of the eigenvalue spectrum 
that allow us to characterize the distribution of incoherent processes in the superoperators,
so that the reduction of these features in our simulated superoperators provides further
evidence for the closer-to-unitary behavior of compensated gates as compared to uncompensated ones.

\subsection{Perturbation Analysis of the Eigenvalue Spectrum}

The eigenvalues of operators in $\mathsf{SU}(2^N)$
lie on a unit circle in the complex plane.
Let $N$ be the number of spins and $U_k$ denote the unitary operator
determined by the RF field in the $k$th frequency interval of the
RF amplitude profile (as shown in \figRef{InhomoProf}).
The eigenvalues of the superoperator obtained by taking
the Kronecker product of $U_k$ with its complex conjugate,
$\bar U_k \otimes U_k$, are the products of the eigenvalues
of $U_k$ with those of $\bar U_k$, yielding $2^N$ eigenvalues
that are equal to unity and $2^{2N-1} - 2^{N-1}$ pairs of complex conjugate eigenvalues.
As a consequence, the weighted sum of the superoperators $\overline{U}_k \otimes U_k$
forces the net superoperator $S$ to have eigenvalues that are symmetric with respect
to the real axis (which is essentially a consequence of the fact that
$S$ preserves Hermiticity) and inside the unit circle.

Estimates of the actual eigenvalues of $S$ will now be
obtained using first-order pertubation theory.
Because the pulses are not perfect even in the absence
of RF field inhomogeneity, we may assume that
the unperturbed eigenvalues are generically non-degenerate.
The unitary operator $U_k$ may be written in exponential form as
\beqn
U_k = e^{-i H_k\, t}
\eeqn
where $H_k$ represents the effective Hamiltonian of the
evolution over the period $t$ for which the pulse is applied.
Defining $H_0$ to be the unperturbed Hamiltonian
(taken to be at the peak of the profile in \figRef{InhomoProf}),
the eigenvalues $\phi_j$ and eigenkets $\mket{\phi_j}$ of $H_{0}$
satisfy the eigenvalue equation
\begin{equation}
U_{0} \mket{\phi_j}=e^{-i\phi_j\,t}\mket{\phi_j}
\end{equation}
where $U_0 = \exp( -i H_0\, t )$.
Note that this implies $\overline{U}_{0} \overline{\mket{\phi_j}}=e^{i\phi_j\,t}\overline{\mket{\phi_j}}$,
where $\overline{\mket{\phi_j}}$ is the complex conjugate of the ket $\mket{\phi_j}$.
The Hamiltonian of a particular $U_k$ can be written as
\begin{equation}
H_k=H_{0} + K_k
\end{equation}
where $K_k$ is the perturbation.
To first order, the new eigenvalues of $H_k$ are 
\begin{equation}
\tilde{\phi}_{j,k}=\phi_j+ \mbra{\phi_j} K_k \mket{\phi_j}
\end{equation}
The corresponding eigenvalues of $U_k$ are then
\begin{equation}
e^{-i\tilde{\phi}_{j,k}\,t}=e^{-i(\phi_j + \mbra{\phi_j} K_k \mket{\phi_j})t}
\end{equation}
Given that $S=\sum_k p_k \overline{U}_k\otimes U_k$, the spectral decomposition of $S$ is
\begin{equation}
S=\sum_k p_k (\sum_m e^{i\tilde{\phi}_{m,k}\,t}\overline{\mket{\phi_m}}\overline{\mbra{\phi_m}} \otimes
\sum_j e^{-i\tilde{\phi}_{j,k}\,t}\mket{\phi_j}\mbra{\phi_j})
\end{equation}
and the eigenvalues of $S$ are then given approximately by
\begin{eqnarray}
\lambda_{jm}&=&\sum_k p_k e^{-i\tilde{\phi}_{j,k}\,t}e^{i\tilde{\phi}_{m,k}\,t}\nonumber\\
\label{eigenvalues}
            &=&\sum_k p_k
            e^{-i(\phi_j-\phi_m)t-i(\mbra{\phi_j}K_k\mket{\phi_j}-
            \overline{\mbra{\phi_m}K_k\mket{\phi_m}})t}\\
                                &=& e^{-i(\phi_j-\phi_m)t}\sum_k p_k
                                e^{-i(\mbra{\phi_j}K_k\mket{\phi_j}-
            \mbra{\phi_m}K_k\mket{\phi_m})t}\nonumber
\end{eqnarray}
where in the last line we have used the fact that the diagonal elements of any Hermitian operator
are real.

\subsection{Eigenvalue Spectra of Uncompensated Pulses }
\begin{figure}
\centerline{\includegraphics[width=14cm,height=6.8cm]{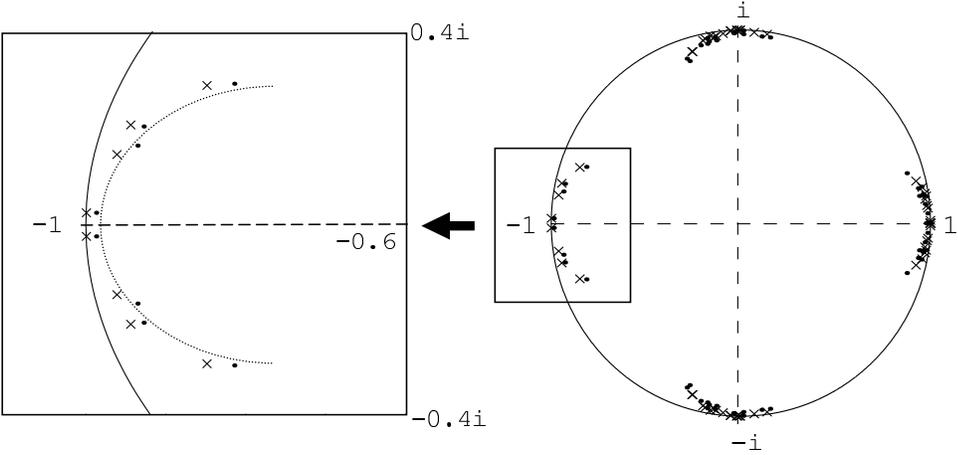}}
\caption{{\bf Eigenvalue spectrum of the simulated superoperator
for a test gate and the radio-frequency inhomogeneity profile shown in \figRef{InhomoProf}}.
The dots are the exact eigenvalues of $S$ while the crosses are
the ones obtained by using the first order perturbation analysis.  The zoom box 
shows some of the detail in the left-hand side of the plot.  The additional 
trend line drawn in the zoomed plot emphasizes the fact that the larger 
the phase shift is, the larger the attenuation is.}
\label{figC1C2}
\end{figure}

We now use the result of the previous section, \eqnRef{eigenvalues}, to calculate
the first-order eigenvalues of an uncompensated $\frac{\pi}{2}$ pulse that rotates
the alanine carbon spins $1$ and $2$ about the $x$-axis.
\figRef{figC1C2} compares the numerically exact
eigenvalues with the results from the approximation.
The dots are the eigenvalues corresponding to the full simulation
of the gate under the influence of the experimental RF inhomogeneity profile.
The crosses are the eigenvalues computed by first-order perturbation theory.
To calculate $\lambda_{jm}$, we first determined $K_k$ using
\begin{eqnarray*}
K_k &=& i\times (\text{log}(U_k)-\text{log}(U_{0}))/t\\
    &=& H_k - H_0
\end{eqnarray*}
where $\text{log}$ is a logarithm of the matrix obtained from the principal branch and then used
\eqnRef{eigenvalues}.

\figRef{figC1C2} also shows an interesting relationship
between the phase shift and the attenuation of the
eigenvalues, i.e. the larger the phase shift, the larger the attenuation.
To get more physical insight of the pattern emerging in \figRef{figC1C2}, we now assume that $K_k$
can be written as $K_k= (\omega_k-\omega_{0})K$. 
Note that we have dropped the
dependence of $K$ with respect to $k$ (i.e.~the "direction" of the effective Hamiltonian is constant
across the sample, which would be expected for a strong RF pulse). 
This would be in fact an exact result for a one spin
system on resonance. The eigenvalue expression then becomes:
\begin{eqnarray}
\lambda_{jm}  &=& e^{-i(\phi_j-\phi_m)t}\sum_k p_k
            e^{-i(\mbra{\phi_j}K_k\mket{\phi_j}-
            \mbra{\phi_m}K_k\mket{\phi_m})t}\\
            &=&  e^{-i(\phi_j-\phi_m)t}\sum_k p_k
            e^{-i\psi_{jm}(\omega_k-\omega_{0})t}
\end{eqnarray}
thereby yielding in the continuous limit
\begin{eqnarray}
\lambda_{jm} &=& e^{-i(\phi_j-\phi_m)t}\int p(\omega_k)
            e^{-i\psi_{jm}(\omega_k-\omega_{0})t} d\omega_k\\
            &=& e^{-i(\phi_j-\phi_m)t}\int p(\Delta\omega)
            e^{-i\psi_{jm}\Delta\omega t} d\Delta\omega
\end{eqnarray}
where $\Delta\omega=\omega_k-\omega_{0}$ and where $\psi_{jm}$ is an unknown real quantity.
We see in this case that to first order the eigenvalue $\lambda_{jm}$ is
just the unperturbed eigenvalue $e^{-i(\phi_j-\phi_m)t}$ times the Fourier transform of
the RF distribution profile evaluated at $\psi_{jm}$. This result demonstrates that the 
probability distribution profile of an incoherent process could be determined in principle
from the eigenvalue structures of an experimental superoperator and a unitary-desired one.
Knowing the form of $K$ would indeed allow one to build the correspondence between
$\lambda_{jm}$ and $\psi_{jm}$, and then to determine $p(\Delta\omega)$ by performing an inverse
Fourier transform. 
This result holds of course when the perturbation is in the first order regime 
and when the unperturbed eigenvalues are non-degenerate.

To gain further understanding about the estimation of the eigenvalues
of $S$, $K_k$ can be expanded  in the basis of products of Pauli
spin matrices and the matrix elements contributing to the perturbation 
theory analysis can be determined. As an example,
let us consider a one-spin system. In this case, $K_k$ can be expanded in terms of
$\sigma_x$, $\sigma_y$ and $\sigma_z$. If we take
$H_0\, t = \frac{\pi}{4}\sigma_x$, then we see that
any operator in the expansion of $K_k$ that anticommutes
with $\sigma_x$ results in zero diagonal matrix elements
in the eigenbasis of $\sigma_x$, because
\begin{eqnarray}
A\sigma_x+\sigma_xA = 0 &\Rightarrow& \mbra{\pm}A\sigma_x+\sigma_xA\mket{\pm}=0\\\nonumber
        &\Rightarrow& \pm 2\mbra{\pm}A\mket{\pm} = 0
\end{eqnarray}
where $\mket{\pm}$ denote the eigenstates of $\sigma_x$,
and $A$ is an operator in the decomposition of $K_k$
that anticommutes with $\sigma_x$.  In this simple example,
only the component along $H_0$ in $K_k$ yields a non-zero contribution in
the eigenvalue calculation given above ($I$ is not present either).
In our three-spin system, if $H_0\,t=\frac{\pi}{4}\sigma_x^1$, it follows that
out of the 64 possible operators in the expansion of $K_k$ only $\sigma_x^1$,
$\sigma_i^2$, $\sigma_j^3$, $\sigma_i^2\sigma_j^3$, $\sigma_x^1\sigma_i^2$,
$\sigma_x^1\sigma_j^3$ and $\sigma_x^1\sigma_i^2\sigma_j^3$
will give a nonzero contribution, where $i,j = x, y$ or $z$.
In general, the number of operators that yield a non-zero
contribution for a $2^n$ by $2^n$ unitary operator $U$ is $2^n-1$, i.e.
the number of diagonal elements minus $1$ (the identity part).
This quick analysis reveals to some extent the slight discrepancy we have
between the first order perturbation theory results and the fully simulated ones.  Due to the nature
of our physical system, the first order perturbation approach takes into
account only a small number of elements in the decomposition of $K_k$.  However, because it
captures the general distribution of the eigenvalue spectrum, we can conclude that a
significant amount of the perturbation is contained in these few operators.

\subsection{Symmetric Inhomogeneity Profile}

\begin{figure}
\centerline{\includegraphics[width=14cm,height=6.8cm]{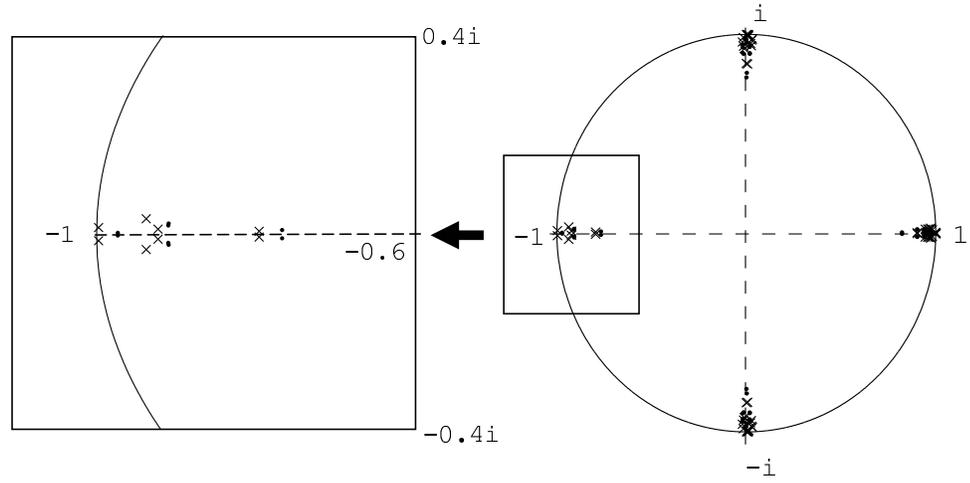}}
\caption{{\bf Eigenvalue spectrum of the simulated superoperator
for the symmetric radio-frequency inhomogeneity profile}. The
dots are the unperturbed eigenvalues and the crosses are the ones computed by using
first order perturbation theory. The symmetry in the distribution mainly results in
some attenuation of the eigenvalues with no phase shift.  The zoom box allows
more detail to be seen for the eigenvalues close to $-i$.}
\label{C1C2sym}
\end{figure}

To test further our analysis and the validity of our assumptions
we calculated the superoperators where a symmetric
inhomogeneity profile is present. We recall the previous formula
\begin{eqnarray}
\lambda_{jm}  &=& e^{-i(\phi_j-\phi_m)t}\sum_k p_k
            e^{-i(\mbra{\phi_j}K_k\mket{\phi_j}-
            \mbra{\phi_m}K_k\mket{\phi_m})t}\\\nonumber
            &=&  e^{-i(\phi_j-\phi_m)t}\sum_k p(\Delta\omega_k)
            e^{-iV_{jm}(\Delta\omega_k)t}
\end{eqnarray}
where $V_{jm}(\Delta\omega_k)=(\mbra{\phi_j}K_k\mket{\phi_j}-\mbra{\phi_m}K_k\mket{\phi_m})$ and
$\Delta\omega_k=\omega_k-\omega_{0}$. Now because $p(\Delta\omega_k)$ is symmetric with respect to $0$, one can
see that the additional assumption of 
$V_{jm}(-\Delta\omega_k)=-V_{jm}(\Delta\omega_k)$ (as we did in the previous
subsection by setting $K_k=\Delta\omega_k K$) leads to the
result that $\lambda_{jm}=e^{-i(\phi_j-\phi_m)t}A_{jm}$ where $A_{jm}$ is a real number smaller
than $1$. In other words, in the case of a symmetric inhomogeneity profile, and provided the above
assumption is valid, the eigenvalues
simply get attenuated, and are not phase shifted. 
As one can see in \figRef{C1C2sym}, for a symmetric profile, the eigenvalues
have practically the same phase as when there is no RF inhomogeneity and are simply attenuated
by the real factor $A_{jm}$. This provides further evidence that
the new eigenvalues are correlated with the Fourier transform of the inhomogeneity profile,
since the Fourier transform of a symmetric real function is real.

\subsection{Eigenvalue Spectra of the Compensated Pulses}
\begin{figure}
\centerline{\includegraphics[width=14cm,height=6.8cm]{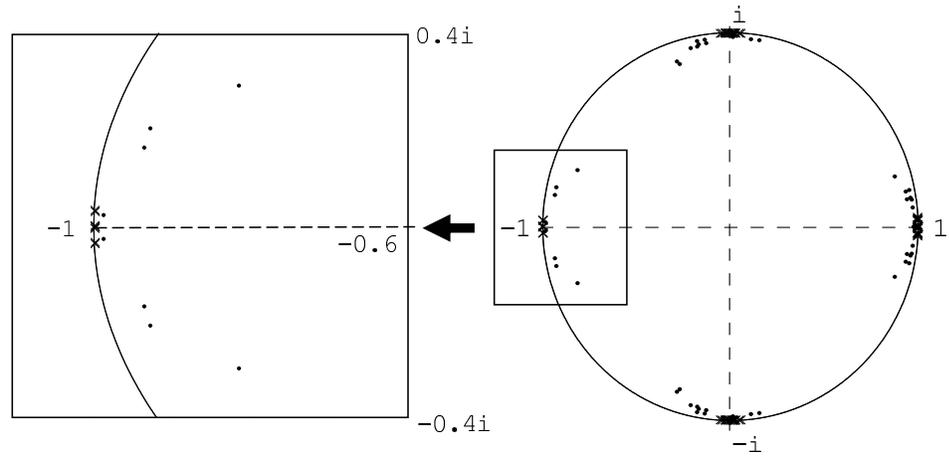}}
\caption{{\bf Eigenvalue spectrum of the simulated superoperators corresponding to a compensated
and uncompensated pulse}. The dots correspond to the uncompensated gate, while
the crosses correspond to the compensated one.  Note that the crosses basically 
lie on the unit circle while the dots are spread inside, confirming the closer-to-unitary
behavior of the compensated gates.}
\label{C1C2Rob}
\end{figure}

As described above, pulses that are compensated for RF inhomogeneity result in an overall
operation closer to being unitary than for uncompensated pulses. In \figRef{C1C2Rob}, we compare the
eigenvalue spectrum of the superoperators corresponding to a compensated and uncompensated
$\frac{\pi}{2}$ pulse
about the $x$ axis on the alanine spins $1$ and $2$.

In general, a unitary superoperator in the Zeeman basis \cite{Havel-Superop} which is trace-preserving
and completely positive must correspond to a unitary process
in Hilbert space, and can be written as $\overline{U}\otimes U$ 
(Boulant \emph{et al.}, in preparation). Since the superoperator of an incoherent
process can be written as a trace-preserving Kraus operator sum,
it must be completely positive and hence is physically reasonable.
Thus the fact that the eigenvalues of the superoperators that were
simulated for the compensated pulses basically lie on the unit circle
provides further evidence for them being very nearly unitary.

\section{Experimental Results}

The experimental tests were carried out on the three carbons of 
$^{13}C$-labeled alanine using a 300 MHz Bruker Avance spectrometer
(see Ref.~\cite{Fortunato-Control} for the internal Hamiltonian's parameters).
The experiments tested compensated and uncompensated versions of seven different 
spin-rotation pulses, which were then used to selectively evolve the two scalar
couplings $J_{12}$ and $J_{23}$ while refocusing the other couplings
\footnote{The waveforms of all the pulses were corrected using a RF feedback procedure,
reducing distortions caused by amplifier nonlinearities
in the experimental setup \cite{Pravia-Thesis}.}.
These pulses were applied to the three input states
\beqn
\rho_{in}=I_j^1+I_j^2+I_j^3, 
\eeqn 
where $j={x,y,z}$. The input states were created using compensated
or uncompensated pulses, depending on the type of gate being tested.
For each gate tested, the average over the three input states 
was calculated and are shown in Tables \ref{SummaryResultsII}
and \ref{PulseResultsII}. Table \ref{SummaryResultsII} shows
the results for the input states, the states obtained by selective
coupling, and an average over all the single-pulse experiments,
while Table \ref{PulseResultsII} shows the specific results
for each single-pulse experiment.

The input and output density matrices were measured
using state tomography and were used in Eqs. \eqn{StateFid}
and \eqn{AttnStateFid} to  evaluate the correlation $C$, attenuation $A$,
and attenuated correlation $C_A$ ($C_A=C\cdot A$).
In all cases, the result of state
tomography on the thermal state $I_z^1+I_z^2+I_z^3$ was used as
the reference for the attenuation ($\rho_{in}$ in  Eq. \eqn{AttnStateFid}).
State tomography \cite{Chuang-StateTomo}
employs read-out pulses to rotate unobservable
elements of the density matrix into observable single-spin transitions.
For the three-spin alanine system used here, eight repetitions
of the experiment,  each with a different readout pulse,
were used to reconstruct the density matrix
\footnote{The eight readout transformations used for the density 
matrix reconstruction were: identity, $\pi/2 ]_y^1$, $\pi/2 ]_y^{12}$, 
$\pi/2 ]_x^{23}$, $\pi/2 ]_y^{3}$, $\pi/2 ]_x^{3}$, 
$\pi/2 ]_x^{1,2,3}$, $\pi/2 ]_y^{1,2,3}$.  All of the 
density matrices dicussed in this section were acquired
using the eight readouts, with the exception of the input states,
which used only the last seven. The last seven readouts alone are also
sufficient, but we included the identity because of its particularly
simple implementation.}.

To determine the actual experimental gate performed, however, 
one requires knowledge about the effect of the gate on a complete 
set of input states so that the full superoperator can be determined, 
a procedure called quantum process tomography
\cite{Chuang-QBlackBox,Childs-ProcTomo,BoulantEtAl}.
To carry out process tomography for a single alanine gate requires
state tomography of the input and output density matrices for a
set of $64$ linearly independent inputs.  Thus full process tomography for 
one gate would involve $2\times64\times8 = 1024$ separate experiments.
While this is certainly possible, it is quite laborious.
For this reason we limited ourselves to performing state tomography on
just the three input states and the corresponding output states described above.
The results do not fully characterize the experimental transformations,
but they provide a reasonable estimate of the quality of the gates.

The most notable difference between the results for compensated and uncompensated
gates is in the severe attenuation caused by RF inhomogeneity on the
uncompensated gates. The average attenuation value $A_{comp.}$ for the
spin rotations is nearly four
times closer to unity than the corresponding value for normal
pulses (i.e.~$(1-A_{uncomp.})/(1-A_{comp.}) = 3.92$). In
addition, $A_{comp.}$ for the $J_{23}$ coupling is more than $10$
times closer to unity than $A_{uncomp.}$, although $A_{comp.}$
for $J_{12}$ is only slightly higher than $A_{uncomp.}$.
The correlations for the compensated rotation gates averaged $0.991$, slightly but
consistently below the average of $0.995$ for the uncompensated gates.  The
attenuated correlations for the compensated gates, however, were higher than the
corresponding values for normal pulses, and the difference was caused
by the sharp differences in the attenuations.  The results confirm
that incorporating RF inhomogeneity into the pulse design can
yield more robust gates and narrows the gap between the
experimental implementation and the simulation results.

\section{Conclusions}
In this report, we explored the capabilities of numerically-determined
strongly-modulating pulses to achieve high gate fidelities in the
presence of a common variety of experimental imperfections.
In particular, we showed that robustness against incoherent errors
such as RF field inhomogeneity can be obtained when knowledge of
these errors is incorporated into the pulse design process.
The basic ideas used in this paper, i.e.~strong
modulation, numerical pulse-design procedures, and the
incorporation of incoherent errors in these procedures,
are likely to find broad applicability in the development
of quantum information processing devices based in many
diverse physical systems. 

\section{Acknowledgements}
This work was supported by ARO, DARPA and NSF. 
Correspondence and requests for materials should be addressed
to D. G. Cory (e-mail: dcory@mit.edu). We thank Zhiying Chen,
Seth Lloyd, and Marcos Saraceno for valuable discussions.

\bibliography{TheBib}

\begin{thebibliography}{31}
\expandafter\ifx\csname natexlab\endcsname\relax\def\natexlab#1{#1}\fi
\expandafter\ifx\csname bibnamefont\endcsname\relax
  \def\bibnamefont#1{#1}\fi
\expandafter\ifx\csname bibfnamefont\endcsname\relax
  \def\bibfnamefont#1{#1}\fi
\expandafter\ifx\csname citenamefont\endcsname\relax
  \def\citenamefont#1{#1}\fi
\expandafter\ifx\csname url\endcsname\relax
  \def\url#1{\texttt{#1}}\fi
\expandafter\ifx\csname urlprefix\endcsname\relax\def\urlprefix{URL }\fi
\providecommand{\bibinfo}[2]{#2}
\providecommand{\eprint}[2][]{\url{#2}}

\bibitem[{\citenamefont{Rabi et~al.}(1939)\citenamefont{Rabi, Millman, Kusch,
  and Zacharias}}]{Rabi}
\bibinfo{author}{\bibfnamefont{I.}~\bibnamefont{Rabi}},
  \bibinfo{author}{\bibfnamefont{S.}~\bibnamefont{Millman}},
  \bibinfo{author}{\bibfnamefont{P.}~\bibnamefont{Kusch}}, \bibnamefont{and}
  \bibinfo{author}{\bibfnamefont{J.}~\bibnamefont{Zacharias}},
  \bibinfo{journal}{Phys. Rev.} \textbf{\bibinfo{volume}{55}},
  \bibinfo{pages}{526} (\bibinfo{year}{1939}).

\bibitem[{\citenamefont{Haeberlen and Waugh}(1968)}]{Waugh}
\bibinfo{author}{\bibfnamefont{U.}~\bibnamefont{Haeberlen}} \bibnamefont{and}
  \bibinfo{author}{\bibfnamefont{J.}~\bibnamefont{Waugh}},
  \bibinfo{journal}{Phys. Rev.} \textbf{\bibinfo{volume}{175}},
  \bibinfo{pages}{453} (\bibinfo{year}{1968}).

\bibitem[{\citenamefont{Haught}(1968)}]{Haught-Control}
\bibinfo{author}{\bibfnamefont{A.}~\bibnamefont{Haught}},
  \bibinfo{journal}{Ann. Rev. Phys. \& Chem.} \textbf{\bibinfo{volume}{19}},
  \bibinfo{pages}{343} (\bibinfo{year}{1968}).

\bibitem[{\citenamefont{Warren et~al.}(1993)\citenamefont{Warren, Rabitz, and
  Dahleh}}]{Warren-Control}
\bibinfo{author}{\bibfnamefont{W.}~\bibnamefont{Warren}},
  \bibinfo{author}{\bibfnamefont{H.}~\bibnamefont{Rabitz}}, \bibnamefont{and}
  \bibinfo{author}{\bibfnamefont{M.}~\bibnamefont{Dahleh}},
  \bibinfo{journal}{Science} \textbf{\bibinfo{volume}{259}},
  \bibinfo{pages}{1581} (\bibinfo{year}{1993}).

\bibitem[{\citenamefont{Warren}(1988)}]{Warren-Review}
\bibinfo{author}{\bibfnamefont{W.}~\bibnamefont{Warren}},
  \bibinfo{journal}{Science} \textbf{\bibinfo{volume}{242}},
  \bibinfo{pages}{878} (\bibinfo{year}{1988}).

\bibitem[{\citenamefont{Gordon and Rice}(1997)}]{GordonR97}
\bibinfo{author}{\bibfnamefont{R.~J.} \bibnamefont{Gordon}} \bibnamefont{and}
  \bibinfo{author}{\bibfnamefont{S.~A.} \bibnamefont{Rice}},
  \bibinfo{journal}{Annu. Rev. Phys. Chem.} \textbf{\bibinfo{volume}{48}},
  \bibinfo{pages}{601} (\bibinfo{year}{1997}).

\bibitem[{\citenamefont{Rice and Shah}(2002)}]{Rice-Active-Control-Review}
\bibinfo{author}{\bibfnamefont{S.}~\bibnamefont{Rice}} \bibnamefont{and}
  \bibinfo{author}{\bibfnamefont{S.}~\bibnamefont{Shah}},
  \bibinfo{journal}{Phys. Chem. Chem. Phys.} \textbf{\bibinfo{volume}{4}},
  \bibinfo{pages}{1683} (\bibinfo{year}{2002}).

\bibitem[{\citenamefont{Ernst et~al.}(1994)\citenamefont{Ernst, Bodenhausen,
  and Wokaun}}]{ErnstBook}
\bibinfo{author}{\bibfnamefont{R.}~\bibnamefont{Ernst}},
  \bibinfo{author}{\bibfnamefont{G.}~\bibnamefont{Bodenhausen}},
  \bibnamefont{and} \bibinfo{author}{\bibfnamefont{A.}~\bibnamefont{Wokaun}},
  \emph{\bibinfo{title}{Principles of Nuclear Magnetic Resonance in One and Two
  Dimensions}} (\bibinfo{publisher}{Oxford University Press},
  \bibinfo{address}{Oxford}, \bibinfo{year}{1994}).

\bibitem[{\citenamefont{Nielsen and Chuang}(2000)}]{NielsenBook}
\bibinfo{author}{\bibfnamefont{M.}~\bibnamefont{Nielsen}} \bibnamefont{and}
  \bibinfo{author}{\bibfnamefont{I.}~\bibnamefont{Chuang}},
  \emph{\bibinfo{title}{Quantum Computation and Quantum Information}}
  (\bibinfo{publisher}{Cambridge University Press},
  \bibinfo{address}{{Cambridge, UK}}, \bibinfo{year}{2000}).

\bibitem[{\citenamefont{Levitt and Freeman}(1979)}]{LevittF79}
\bibinfo{author}{\bibfnamefont{M.}~\bibnamefont{Levitt}} \bibnamefont{and}
  \bibinfo{author}{\bibfnamefont{R.}~\bibnamefont{Freeman}},
  \bibinfo{journal}{J. Magn. Reson.} \textbf{\bibinfo{volume}{33}},
  \bibinfo{pages}{473} (\bibinfo{year}{1979}).

\bibitem[{\citenamefont{Levitt}(1982)}]{Levitt82}
\bibinfo{author}{\bibfnamefont{M.}~\bibnamefont{Levitt}}, \bibinfo{journal}{J.
  Magn. Reson.} \textbf{\bibinfo{volume}{48}}, \bibinfo{pages}{234}
  (\bibinfo{year}{1982}).

\bibitem[{\citenamefont{Tycko}(1983)}]{Tycko83}
\bibinfo{author}{\bibfnamefont{R.}~\bibnamefont{Tycko}},
  \bibinfo{journal}{Phys. Rev. Lett.} \textbf{\bibinfo{volume}{51}},
  \bibinfo{pages}{775} (\bibinfo{year}{1983}).

\bibitem[{\citenamefont{Shaka and Freeman}(1983)}]{Shaka-DualComp}
\bibinfo{author}{\bibfnamefont{A.}~\bibnamefont{Shaka}} \bibnamefont{and}
  \bibinfo{author}{\bibfnamefont{R.}~\bibnamefont{Freeman}},
  \bibinfo{journal}{J. Magn. Reson,} \textbf{\bibinfo{volume}{55}},
  \bibinfo{pages}{487} (\bibinfo{year}{1983}).

\bibitem[{\citenamefont{Levitt}(1986)}]{Levitt86}
\bibinfo{author}{\bibfnamefont{M.}~\bibnamefont{Levitt}},
  \bibinfo{journal}{Prog. Nucl. Magn. Reson. Spect.}
  \textbf{\bibinfo{volume}{18}}, \bibinfo{pages}{61} (\bibinfo{year}{1986}).

\bibitem[{\citenamefont{Baum et~al.}(1985)\citenamefont{Baum, Tycko, and
  Pines}}]{BaumTP85}
\bibinfo{author}{\bibfnamefont{J.}~\bibnamefont{Baum}},
  \bibinfo{author}{\bibfnamefont{R.}~\bibnamefont{Tycko}}, \bibnamefont{and}
  \bibinfo{author}{\bibfnamefont{A.}~\bibnamefont{Pines}},
  \bibinfo{journal}{Phys. Rev. A} \textbf{\bibinfo{volume}{32}},
  \bibinfo{pages}{3435} (\bibinfo{year}{1985}).

\bibitem[{\citenamefont{Silver et~al.}(1985)\citenamefont{Silver, Joseph, and
  Hoult}}]{SilverJH85}
\bibinfo{author}{\bibfnamefont{M.~S.} \bibnamefont{Silver}},
  \bibinfo{author}{\bibfnamefont{R.~I.} \bibnamefont{Joseph}},
  \bibnamefont{and} \bibinfo{author}{\bibfnamefont{D.~I.} \bibnamefont{Hoult}},
  \bibinfo{journal}{Phys. Rev. A} \textbf{\bibinfo{volume}{31}},
  \bibinfo{pages}{2753} (\bibinfo{year}{1985}).

\bibitem[{\citenamefont{Cory}(1993)}]{Cory-RFSelection}
\bibinfo{author}{\bibfnamefont{D.}~\bibnamefont{Cory}}, \bibinfo{journal}{J.
  Magn. Reson} \textbf{\bibinfo{volume}{103}}, \bibinfo{pages}{23}
  (\bibinfo{year}{1993}).

\bibitem[{\citenamefont{Cummins et~al.}(2003)\citenamefont{Cummins, Llewellyn,
  and Jones}}]{JonesRobust}
\bibinfo{author}{\bibfnamefont{H.}~\bibnamefont{Cummins}},
  \bibinfo{author}{\bibfnamefont{G.}~\bibnamefont{Llewellyn}},
  \bibnamefont{and} \bibinfo{author}{\bibfnamefont{J.}~\bibnamefont{Jones}},
  \bibinfo{journal}{Phys. Rev. A} \textbf{\bibinfo{volume}{67}},
  \bibinfo{pages}{042308} (\bibinfo{year}{2003}).

\bibitem[{\citenamefont{Fortunato
  et~al.}(2002{\natexlab{a}})\citenamefont{Fortunato, Pravia, Boulant,
  Teklemariam, Havel, and Cory}}]{Fortunato-Control}
\bibinfo{author}{\bibfnamefont{E.}~\bibnamefont{Fortunato}},
  \bibinfo{author}{\bibfnamefont{M.}~\bibnamefont{Pravia}},
  \bibinfo{author}{\bibfnamefont{N.}~\bibnamefont{Boulant}},
  \bibinfo{author}{\bibfnamefont{G.}~\bibnamefont{Teklemariam}},
  \bibinfo{author}{\bibfnamefont{T.}~\bibnamefont{Havel}}, \bibnamefont{and}
  \bibinfo{author}{\bibfnamefont{D.}~\bibnamefont{Cory}}, \bibinfo{journal}{J.
  Chem. Phys.} \textbf{\bibinfo{volume}{116}}, \bibinfo{pages}{7599}
  (\bibinfo{year}{2002}{\natexlab{a}}).

\bibitem[{\citenamefont{Cory et~al.}(1997)\citenamefont{Cory, Fahmy, and
  Havel}}]{Cory-PP}
\bibinfo{author}{\bibfnamefont{D.}~\bibnamefont{Cory}},
  \bibinfo{author}{\bibfnamefont{A.}~\bibnamefont{Fahmy}}, \bibnamefont{and}
  \bibinfo{author}{\bibfnamefont{T.}~\bibnamefont{Havel}},
  \bibinfo{journal}{Proc. Natl. Acad. Sci.} \textbf{\bibinfo{volume}{94}},
  \bibinfo{pages}{1634} (\bibinfo{year}{1997}).

\bibitem[{\citenamefont{Gershenfeld and Chuang}(1997)}]{Chuang-PP}
\bibinfo{author}{\bibfnamefont{N.}~\bibnamefont{Gershenfeld}} \bibnamefont{and}
  \bibinfo{author}{\bibfnamefont{I.}~\bibnamefont{Chuang}},
  \bibinfo{journal}{Science} \textbf{\bibinfo{volume}{275}},
  \bibinfo{pages}{350} (\bibinfo{year}{1997}).

\bibitem[{\citenamefont{Jones}(2001)}]{Jones-NMRQIP}
\bibinfo{author}{\bibfnamefont{J.}~\bibnamefont{Jones}},
  \bibinfo{journal}{Prog. in {NMR} Spect.} \textbf{\bibinfo{volume}{38}},
  \bibinfo{pages}{325} (\bibinfo{year}{2001}).

\bibitem[{\citenamefont{Cory et~al.}(2000)\citenamefont{Cory, Laflamme, Knill,
  Viola, Havel, Boulant, Boutis, Fortunato, S, R et~al.}}]{Cory-Overview}
\bibinfo{author}{\bibfnamefont{D.}~\bibnamefont{Cory}},
  \bibinfo{author}{\bibfnamefont{R.}~\bibnamefont{Laflamme}},
  \bibinfo{author}{\bibfnamefont{E.}~\bibnamefont{Knill}},
  \bibinfo{author}{\bibfnamefont{L.}~\bibnamefont{Viola}},
  \bibinfo{author}{\bibfnamefont{T.}~\bibnamefont{Havel}},
  \bibinfo{author}{\bibfnamefont{N.}~\bibnamefont{Boulant}},
  \bibinfo{author}{\bibfnamefont{G.}~\bibnamefont{Boutis}},
  \bibinfo{author}{\bibfnamefont{E.}~\bibnamefont{Fortunato}},
  \bibinfo{author}{\bibfnamefont{S.~L.} \bibnamefont{S}},
  \bibinfo{author}{\bibfnamefont{R.~M.} \bibnamefont{R}}, \bibnamefont{et~al.},
  \bibinfo{journal}{Prog. Phys.} \textbf{\bibinfo{volume}{48}},
  \bibinfo{pages}{875} (\bibinfo{year}{2000}).

\bibitem[{\citenamefont{Havel}(2003)}]{Havel-Superop}
\bibinfo{author}{\bibfnamefont{T.}~\bibnamefont{Havel}}, \bibinfo{journal}{J.
  Math. Phys.} \textbf{\bibinfo{volume}{44}}, \bibinfo{pages}{534}
  (\bibinfo{year}{2003}).

\bibitem[{\citenamefont{Schumacher}(1996)}]{Schumacher-Channels}
\bibinfo{author}{\bibfnamefont{B.}~\bibnamefont{Schumacher}},
  \bibinfo{journal}{Phys. Rev. A} \textbf{\bibinfo{volume}{54}},
  \bibinfo{pages}{2614} (\bibinfo{year}{1996}).

\bibitem[{\citenamefont{Fortunato
  et~al.}(2002{\natexlab{b}})\citenamefont{Fortunato, Viola, Hodges,
  Teklemariam, and Cory}}]{Fortunato-DFS}
\bibinfo{author}{\bibfnamefont{E.}~\bibnamefont{Fortunato}},
  \bibinfo{author}{\bibfnamefont{L.}~\bibnamefont{Viola}},
  \bibinfo{author}{\bibfnamefont{J.}~\bibnamefont{Hodges}},
  \bibinfo{author}{\bibfnamefont{G.}~\bibnamefont{Teklemariam}},
  \bibnamefont{and} \bibinfo{author}{\bibfnamefont{D.}~\bibnamefont{Cory}},
  \bibinfo{journal}{New J. Phys.} \textbf{\bibinfo{volume}{4}},
  \bibinfo{pages}{5.1} (\bibinfo{year}{2002}{\natexlab{b}}).

\bibitem[{\citenamefont{Chuang et~al.}(1998)\citenamefont{Chuang, Gershenfeld,
  Kubinec, and Leung}}]{Chuang-StateTomo}
\bibinfo{author}{\bibfnamefont{I.}~\bibnamefont{Chuang}},
  \bibinfo{author}{\bibfnamefont{N.}~\bibnamefont{Gershenfeld}},
  \bibinfo{author}{\bibfnamefont{M.}~\bibnamefont{Kubinec}}, \bibnamefont{and}
  \bibinfo{author}{\bibfnamefont{D.}~\bibnamefont{Leung}},
  \bibinfo{journal}{Proc. R. Soc. Lond. A} \textbf{\bibinfo{volume}{454}},
  \bibinfo{pages}{447} (\bibinfo{year}{1998}).

\bibitem[{\citenamefont{Chuang and Nielsen}(1997)}]{Chuang-QBlackBox}
\bibinfo{author}{\bibfnamefont{I.}~\bibnamefont{Chuang}} \bibnamefont{and}
  \bibinfo{author}{\bibfnamefont{M.}~\bibnamefont{Nielsen}},
  \bibinfo{journal}{J. Mod. Opt.} \textbf{\bibinfo{volume}{44}},
  \bibinfo{pages}{2455} (\bibinfo{year}{1997}).

\bibitem[{\citenamefont{Childs et~al.}(2001)\citenamefont{Childs, Chuang, and
  Leung}}]{Childs-ProcTomo}
\bibinfo{author}{\bibfnamefont{A.}~\bibnamefont{Childs}},
  \bibinfo{author}{\bibfnamefont{I.}~\bibnamefont{Chuang}}, \bibnamefont{and}
  \bibinfo{author}{\bibfnamefont{D.}~\bibnamefont{Leung}},
  \bibinfo{journal}{Phys. Rev. A} \textbf{\bibinfo{volume}{64}},
  \bibinfo{pages}{012314/1} (\bibinfo{year}{2001}).

\bibitem[{\citenamefont{Boulant et~al.}(2003)\citenamefont{Boulant, Havel,
  Pravia, and Cory}}]{BoulantEtAl}
\bibinfo{author}{\bibfnamefont{N.}~\bibnamefont{Boulant}},
  \bibinfo{author}{\bibfnamefont{T.~F.} \bibnamefont{Havel}},
  \bibinfo{author}{\bibfnamefont{M.~A.} \bibnamefont{Pravia}},
  \bibnamefont{and} \bibinfo{author}{\bibfnamefont{D.~G.} \bibnamefont{Cory}},
  \bibinfo{journal}{Phys. Rev. A} \textbf{\bibinfo{volume}{67}},
  \bibinfo{pages}{042322} (\bibinfo{year}{2003}).

\bibitem[{\citenamefont{Pravia}(2002)}]{Pravia-Thesis}
\bibinfo{author}{\bibfnamefont{M.}~\bibnamefont{Pravia}}, Ph.D. thesis,
  \bibinfo{school}{Massachusetts Institute of Technology}
  (\bibinfo{year}{2002}).

\end{thebibliography}
\bibliographystyle{apsrev}

\clearpage
\newpage

\section{Tables and Figures}

\begin{table} [htb] \centering 
\begin{tabular}{ccccc} 
Control\hspace{1cm} & Input\hspace{1cm}  & Average over Seven  & Selective & Selective \\
Metrics\hspace{1cm} & State\hspace{1cm}  & Spin-Rotation Gates &$J_{12}$ Gate & $J_{23}$ Gate \\[6pt]
\hline \rule{0ex}{3ex}
$\langle C^{comp.} \rangle $\hspace{1cm} &  0.993\hspace{1cm} &  \hspace{1cm}0.991\hspace{1cm} &  \hspace{1cm}0.986\hspace{1cm} &  0.990\\
$\langle C^{uncomp.} \rangle$  \hspace{1cm} &  0.995\hspace{1cm} &  \hspace{1cm}0.995\hspace{1cm} &  \hspace{1cm}0.975\hspace{1cm} &  0.988\\[6pt]
\hline \rule{0ex}{3ex}
$\langle A^{comp.} \rangle$\hspace{1cm} &  0.998\hspace{1cm} &  \hspace{1cm}0.987\hspace{1cm} &  \hspace{1cm}0.954\hspace{1cm} &  0.994\\
$\langle A^{uncomp.} \rangle$\hspace{1cm} &  0.970\hspace{1cm} &  \hspace{1cm}0.949\hspace{1cm} &  \hspace{1cm}0.951\hspace{1cm} &  0.930\\[6pt]
\hline \rule{0ex}{3ex}
$\langle C^{comp.}_A \rangle$\hspace{1cm} &  0.991\hspace{1cm} &  \hspace{1cm}0.979\hspace{1cm} &  \hspace{1cm}0.941\hspace{1cm} &  0.984\\
$\langle C^{uncomp.}_A \rangle$\hspace{1cm} &  0.965\hspace{1cm} &  \hspace{1cm}0.944\hspace{1cm} &  \hspace{1cm}0.927\hspace{1cm} &  0.919\\[6pt]
\hline
\end{tabular}
\caption{{\bf Summary of experimental results}.  
The metrics $C$, $A$, and $C_A$, refer
to the correlation, attenuation, and attenuated correlation ($C_A=C\cdot A$).
The superscripts specify whether the pulses employed were compensated 
or uncompensated for RF inhomogeneity, while the angle brackets denote 
that the reported quantities are means over the three input states tested 
for each transformation.  In the case of the spin-rotation values, the 
quantity reported is the average of all the spin-rotation results.
}  
\label{SummaryResultsII}
\end{table}

\begin{table} [htb] \centering 
\begin{tabular}{cccccccc} 
Metrics\hspace{1cm} & $\pi/2]_x^1$\hspace{1cm} & $\pi / 2]_x^3$\hspace{1cm} & $\pi / 2]_x^{12}$\hspace{1cm} & $\pi / 2]_x^{23}$\hspace{1cm} & $\pi / 2]_x^{123}$\hspace{1cm} & $\pi]_x^{12}$\hspace{1cm} & $\pi]_x^{23}$ \\[6pt]
\hline \rule{0ex}{3ex}
$\langle C^{comp.} \rangle$\hspace{1cm} &  0.991\hspace{1cm} &  0.992\hspace{1cm} &  0.991\hspace{1cm} &  0.986\hspace{1cm} &  0.994\hspace{1cm} &  0.993\hspace{1cm} &  0.992\\
$\langle C^{uncomp.} \rangle$\hspace{1cm} &  0.996\hspace{1cm} &  0.994\hspace{1cm} &  0.996\hspace{1cm} &  0.994\hspace{1cm} &  0.996\hspace{1cm} &  0.995\hspace{1cm} &  0.995\\[6pt]
\hline \rule{0ex}{3ex}
$\langle A^{comp.} \rangle$\hspace{1cm} &  0.990\hspace{1cm} &  0.989\hspace{1cm} &  0.987\hspace{1cm} &  0.988\hspace{1cm} &  0.986\hspace{1cm} &  0.984\hspace{1cm} &  0.986\\
$\langle A^{uncomp.} \rangle$\hspace{1cm} &  0.954\hspace{1cm} &  0.953\hspace{1cm} &  0.948\hspace{1cm} &  0.951\hspace{1cm} &  0.942\hspace{1cm} &  0.952\hspace{1cm} &  0.941\\[6pt]
\hline \rule{0ex}{3ex}
$\langle C^{comp.}_A \rangle$\hspace{1cm} &  0.981\hspace{1cm} &  0.981\hspace{1cm} &  0.978\hspace{1cm} &  0.974\hspace{1cm} &  0.980\hspace{1cm} &  0.977\hspace{1cm} &  0.979\\
$\langle C^{uncomp.}_A \rangle$\hspace{1cm} &  0.950\hspace{1cm} &  0.948\hspace{1cm} &  0.944\hspace{1cm} &  0.945\hspace{1cm} &  0.938\hspace{1cm} &  0.947\hspace{1cm} &  0.936\\[6pt]
\hline \rule{0ex}{3ex}
\end{tabular}
\caption{{\bf Experimental results of spin-rotation gates}.
The metrics $C$, $A$, and $C_A$, refer to the correlation, attenuation, 
and attenuated correlation ($C_A=C\cdot A$).
The superscripts specify whether the pulses employed were compensated 
or uncompensated for RF inhomogeneity, while the angle brackets denote 
that the reported quantities are means over the three input states tested 
for each transformation.  The spin-rotation pulses tested were $\pi/2$ 
and $\pi$ rotations of the carbon spins denoted in the superscript. }
\label{PulseResultsII}
\end{table}

\clearpage

\end{document}